# Excitonic Effects in the Optical Spectra of Graphene Nanoribbons

*Li Yang, Marvin L. Cohen, and Steven G. Louie*

Department of Physics, University of California at Berkeley, California 94720, and Materials Sciences Division, Lawrence Berkeley National Laboratory, Berkeley, California 94720

(Abstract)

*We present a first-principles calculation of the optical properties of armchair-edged graphene nanoribbons (AGNRs) with many-electron effects included. The reduced dimensionality of the AGNRs gives rise to an enhanced electron-hole binding energy for both bright and dark exciton states (0.8-1.4 eV for GNRs with width w ~ 1.2 nm) and dramatically changes the optical spectra owing to a near complete transfer of oscillator strength to the exciton states from the continuum transitions. The characteristics of the excitons of the three distinct families of AGNRs are compared and discussed. The enhanced excitonic effects found here are expected to be of importance in optoelectronic applications of graphene-based nanostructures.*

The successful fabrication and measurement of graphene [1-3], a single atomic layer of graphite, have resulted in intensive research on graphene-based structures because of fundamental physics interests and promising applications [4-13]. A strip of graphene of nanometers in width can be obtained, e.g., by etching or patterning. These structures are referred to as graphene nanoribbons (GNRs). Recent advances in fabricating and characterizing stable GNRs [14,15] provide opportunities to explore the various remarkable properties of these systems.

A recent study [16] has shown that the quasiparticle band gap of armchair-edged GNRs (AGNRs) is in the interesting energy range of 1-3 eV for 2-1 nm wide GNRs and is therefore promising for electronics applications. However, the optical properties of the GNRs have not been explored going beyond the independent-particle picture [17], yet many-electron effects are expected to play an important role in determining GNR optical properties. Recent *ab initio* calculations, as well as experiments, have shown that excitonic effects dramatically alter the optical response of one-dimensional structures, such as carbon and BN nanotubes and Si nanowires [18-23]. Hence, in this work, we carry out a first-principles calculation, including the relevant many-electron (quasiparticle self-energy and electron-hole interaction) effects, to study the optical properties of AGNRs.

Following the approach of Rolfling and Louie [24], we first obtain the electronic ground state using density functional theory (DFT) within the local density approximation (LDA). Next, the quasiparticle energies are calculated within the GW approximation to the electron self-energy [25]. We then compute the photo-excited states by solving the Bethe-Salpeter equation (BSE) of the two-particle Green's function [24]. The electron-hole excited state is represented by the expansion

$$|S\rangle = \sum_k \sum_v^{hole} \sum_c^{elec} A_{vck}^S |vck\rangle.$$

The exciton amplitude $A_{vck}^S$ is obtained by solving the BSE [24,26]:

$$(E_{ck} - E_{vk})A_{vck}^S + \sum_{k'v'c'} \langle vck|K^{eh}|v'c'k'\rangle A_{v'c'k'}^S = \Omega^S A_{vck}^S,$$

where $K^{eh}$ is the electron-hole interaction kernel, $E_{ck}$ and $E_{vk}$ are quasiparticle energies for the conduction band state $|ck\rangle$ and valence band state $|vk\rangle$, respectively, and $\Omega^S$ is the energy of the excited state. In this formalism, the electron-hole amplitude in real space (or the wavefunction of exciton) is given by:

$$\Phi_S(\vec{x}_e, \vec{x}_h) = \sum_k^{hole} \sum_v^{elec} \sum_c A^S_{vck} \phi_{ck}(\vec{x}_e) \phi^*_{vk}(\vec{x}_h).$$

Finally, the absorption spectra are obtained by calculating the imaginary part of the dielectric function as described in Ref [24].

The following notation is used in describing an AGNR (since an AGNR may be specified by the number of dimer lines along the ribbon forming the width): N-AGNR is an AGNR with N dimer lines along the ribbon. In this study, we consider three N-AGNRs (N=10, 11, and 12), which cover the distinct three families (N=3p+1, 3p, and 3p-1, where p is an integer) of AGNRs [13], and their widths vary from 1.1 nm to 1.4 nm. The dangling σ-bonds at the edges are passivated by hydrogen atoms. Norm-conserving pseudopotentials [27] and a plane-wave basis are used, and the k-grid is sampled uniformly along the one-dimensional Brillouin zone. To assure that the quasiparticle energies are converged to within 0.1 eV, a 1x1x32 k-point sampling is necessary. As concluded from previous studies [24], the k-point sampling in solving the BSE should be much denser than that for the GW quasiparticle energy calculation. Hence, the electron-hole interaction kernel $K^{eh}$ [24] is first computed on a coarse *k* grid (1x1x32), then interpolated on a fine grid (1x1x256) to get converged results. For the optical absorption spectrum calculation, 8 valence bands and 8 conduction bands are included. Since the supercell method is used in these calculations, a rectangular-shape truncated Coulomb interaction is applied to eliminate the image effect between adjacent supercells to mimic isolated GNRs [16,19,28,29].

We first discuss the 10-AGNR since the other two AGNRs studied have similar behavior. Figure 1 shows the quasiparticle energy corrections for the 10-AGNR to its LDA Kohn-Sham eigenvalues. The quasiparticle energy correction increases the LDA band gap from 1.3 eV to 3.2 eV. This is a much larger change than those found for bulk graphite or diamond [25]. This large correction is a consequence of the enhanced Coulomb interaction effects for reduced dimensions. In addition, the quasiparticle energy corrections show a complicated band and energy dependence so that there is no simple "scissor rule" to obtain the quasiparticle band structure. The corrections depend on the character of the electronic states. For example, the quasiparticle energy corrections to the nearly-free-electron (NFE) states, which are states loosely bound to the graphene sheet, are significantly smaller than those of the π states. The σ bands located in the valence bands also behave differently. As a result, we have to correct each type of electronic state to obtain the correct quasiparticle band structure.

Figure 2 shows the calculated quasiparticle band structure and the optical spectra of the 10-AGNR. Because of a strong depolarization effect, we show here only the optical absorption for the case with the light polarization vector along the direction of the ribbon length. Optical transitions that give rise to the independent-particle interband optical absorption peaks (cyan colored curves in Figure 2b) are marked with arrows on the band structure in Figure 2a. The interband transitions and the bright excitons are labeled as follows: $E^{nm}$ denotes the interband transition from the n-th valence band to the m-th conduction band (counting the bands from the gap); $E_i^{nm}$ denotes the i-th exciton state associated with the interband transition $E^{nm}$.

As shown in Figure 2b, the first photo-absorption peak of the 10-AGNR appears at an excitation energy of 1.8 eV as a result of a bound exciton with a 1.4 eV binding energy. There are two reasons for such a large binding energy in the AGNRs: 1) quantum confinement which modifies the electron and hole wave functions and density of states; 2) reduced screening owing to a larger quasiparticle band gap and a finite ribbon width surrounded by vacuum. We note that quantum confinement increases the quasiparticle energy gap, while the electron-hole interaction results in large binding energies of excitons. The two effects are both significant but affect the excitation energy in opposite ways. Similar behavior has been observed in the optical response of carbon and BN nanotubes and Si nanowires [18-23]. Therefore, the first absorption peak is at 1.8 eV for the 10-AGNR with many-electron effects included. This is only around 0.5 eV higher than that of the LDA interband transition value, although the nature of the excited state is completely different.

In Figure 2, the energies for the onsets of interband transitions $E^{11}$ and $E^{22}$ are almost degenerated. However, the dispersions of the energy bands giving rise to these transitions are different. Hence, the effective masses of the electron and hole associated with the transitions $E^{11}$ and $E^{22}$ are different as seen in Figure 2a. As a result, the degeneracy in the excitation energies is broken when the electron-hole interaction is included because of a difference in binding energies for the exciton states. We also note that the electron-hole interaction reverses the apparent relative intensity of the first and second prominent ($E_1^{11}$ and $E_1^{22}$) optical peaks for the 10-AGNR.

There are also dark spin-singlet excitons existing around the absorption peaks of $E_1^{11}$ and $E_1^{22}$ in the 10-AGNR. For example, the exciton $D_1$ in Figure 2b arising from the mixture of interband transitions $E^{12}$ and $E^{21}$ is a dark exciton for light polarization vector along the ribbon length. Another dark exciton labeled with $D_2$ in Figure 2b has a higher energy and is the "excited-states" of $D_1$. These dark excitons provide channels for non-radiative decay processes and affect the ribbon's luminescence. On the other hand, with a different light polarization direction or modifications, such as changing the excitation process, these dark excitons may be excited and thus can result in interesting new phenomena.

The optical spectra of the 11- and 12-AGNRs are presented in Figure 3, showing similarly strong exciton features. Comparing the binding energy of the first bright exciton in 10-, 11- and 12-AGNRs, we find that it is sensitive to the band gap of the AGNR. The 10-AGNR has the largest band gap and, therefore, a weakest screening of the Coulomb interaction between electron and hole among the three AGNRs studied yielding in the largest exciton binding energy.

In general, we may classify excitons in semiconductors into: a) bound excitons with energies below the two-particle continuum and b) resonant excitons with energies above the two-particle continuum. We find both types of excitons in the AGNRs, and they both give rise to prominent peaks in the optical spectra. For the 10-AGNR in Figure 2b, $E_1^{11}$, $E_2^{11}$ and $E_1^{22}$ are bound excitons, and $E_1^{36}$ and $E_1^{28}$ is a resonant exciton. Figure 4 shows the real-space, electron-hole probability distribution for three specific excitons of the 10-AGNR: $E_1^{11}$, $E_1^{22}$ and $E_1^{36}$, with the position of the hole fixed on top of a π orbital and marked with the black spot. The character of these excitons reflects the corresponding π electron states. For example, the electron amplitude of the exciton $E_1^{11}$ shown in Figure 4a is that of the first conduction band modulated by the exciton coefficients $A_{vck}^S$. Because of the different wavefunction character between the first and second conduction bands, the electron amplitude of the excitons $E_1^{11}$ and $E_1^{22}$ plotted in Figure 4a and 4b show different fine structures although they have the similar extent. Quantum confinement (i.e., the 6th conduction band is a higher-order subband state in the transverse ribbon direction) results in a nodal line in the electron distribution of the exciton $E_1^{36}$ along the ribbon as shown in Figure 4c.

To understand the real-space localization of excitons in more detail, a more quantitative presentation appears in Figure 5. The projected electron density,

$$\left|\Phi_S(z_e, \vec{x}_h = 0)\right|^2 = \iint dx_e dy_e \left|\Phi_S(\vec{x}_e, \vec{x}_h = 0)\right|^2$$

is plotted along the ribbon length after integrating the electron coordinates in the perpendicular directions keeping the hole fixed on top of a π orbital that has the maximum density for the hole states. Within the same series of excitons, the "excited-state" of a specific bright exciton (such as $E_2^{11}$) has a significantly larger extent than the "ground-state" exciton (such as $E_1^{11}$) as shown in Figure 5a and 5b. Figure 5a, 5c and 5d also make a comparison of the extent of the first bright excitons in 10-, 11- and 12-AGNR. The extents of these excitons are closely related to their corresponding binding energies.

In conclusion, we have performed a detailed first-principles study of the effects of self-energy and electron-hole interaction in the optical response of AGNRs. The quasiparticle energy corrections are found to be subband-dependent so that a simple "scissor rule" shift is inappropriate to obtain an accurate quasiparticle band structure for this system. As in nanotubes, because of reduced dimensionality, excitonic effects are dominant in the optical spectrum of the GNRs. The binding energy of excitons in AGNRs with width around 1.2 nm is found to be 0.8-1.4 eV. In addition, dark excitons are found in the AGNRs, and this is also of importance to the photophysics of these materials. The characters of the bound and resonant excitons in the 10-, 11- and 12-AGNRs studied are compared and explained.

*Note added*: During preparation of this manuscript we become aware of a related theoretical study by D. Prezzi, et al. [30] on the 8-, 9- and 10-AGNRs, which has obtained similar trends to those found in the present work on the 10-, 11- and 12-AGNRs.

We thank C.-H. Park, D. Prendergast and Y.-W. Son for discussions. This research is supported by NSF Grant No. DMR04-39768 and by the Director, Office of Science, Office of Basic Energy under Contract No.DE-AC02-05CH11231. Computational resources have been provided by Datastar at the San Diego Supercomputer Center.

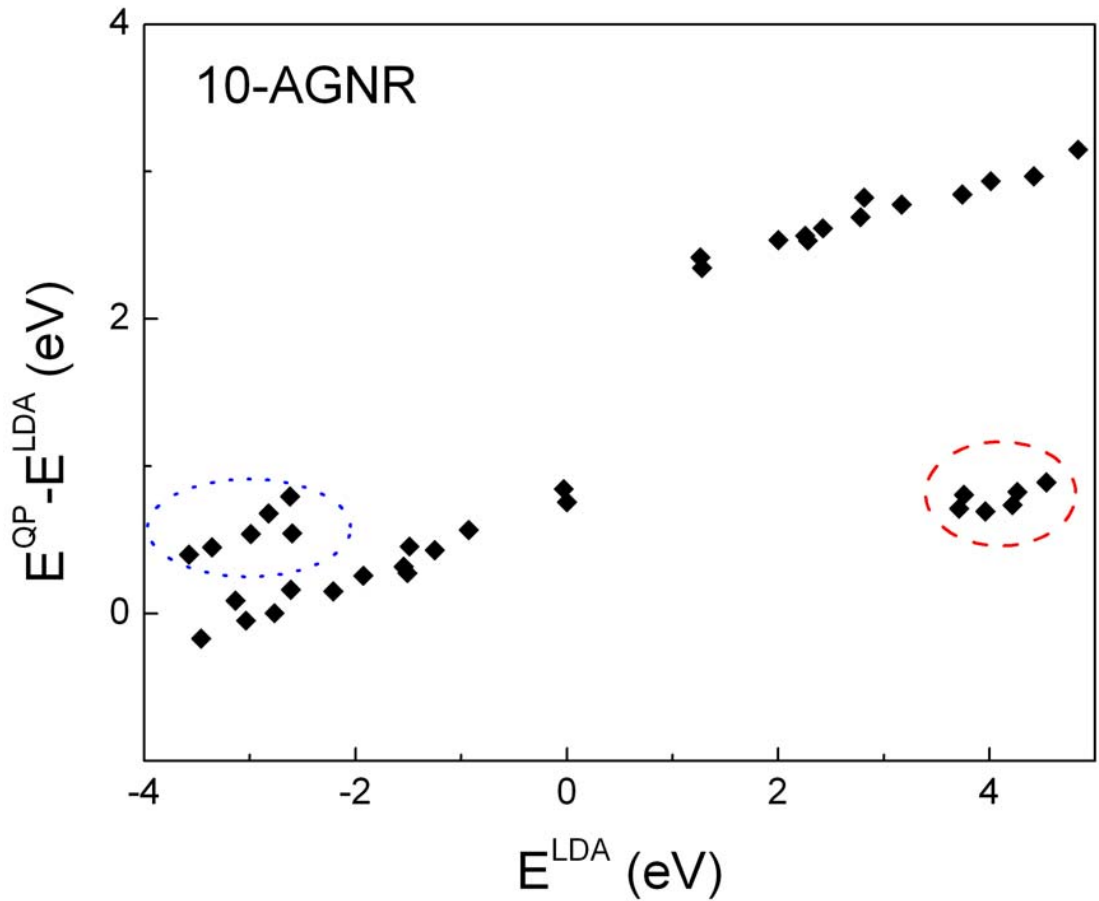

**Figure 1**. Differences between the GW quasiparticle energy and the LDA Kohn-Sham eigenvalue of a 10-AGNR. The top of the LDA valence band is set to be zero. The corrections to the NFE state are marked by a dashed red circle, and the corrections to the σ band are marked by a dotted blue circle.

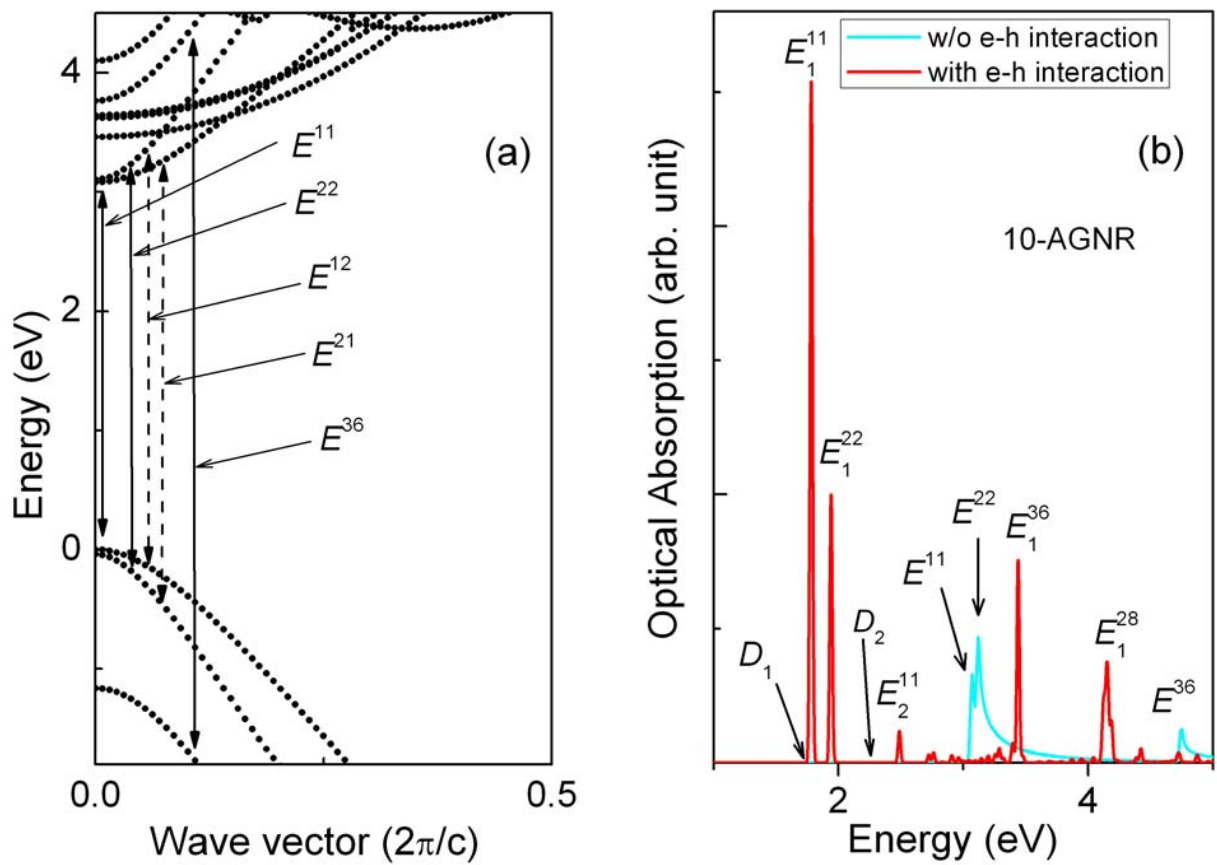

**Figure 2**. Quasiparticle band structure and optical spectra of the 10-AGNR. The top of valence band is set at zero. *c* is the lattice constant along the ribbon length. The spectra are broadened with a gaussian smearing of width of 0.01 eV

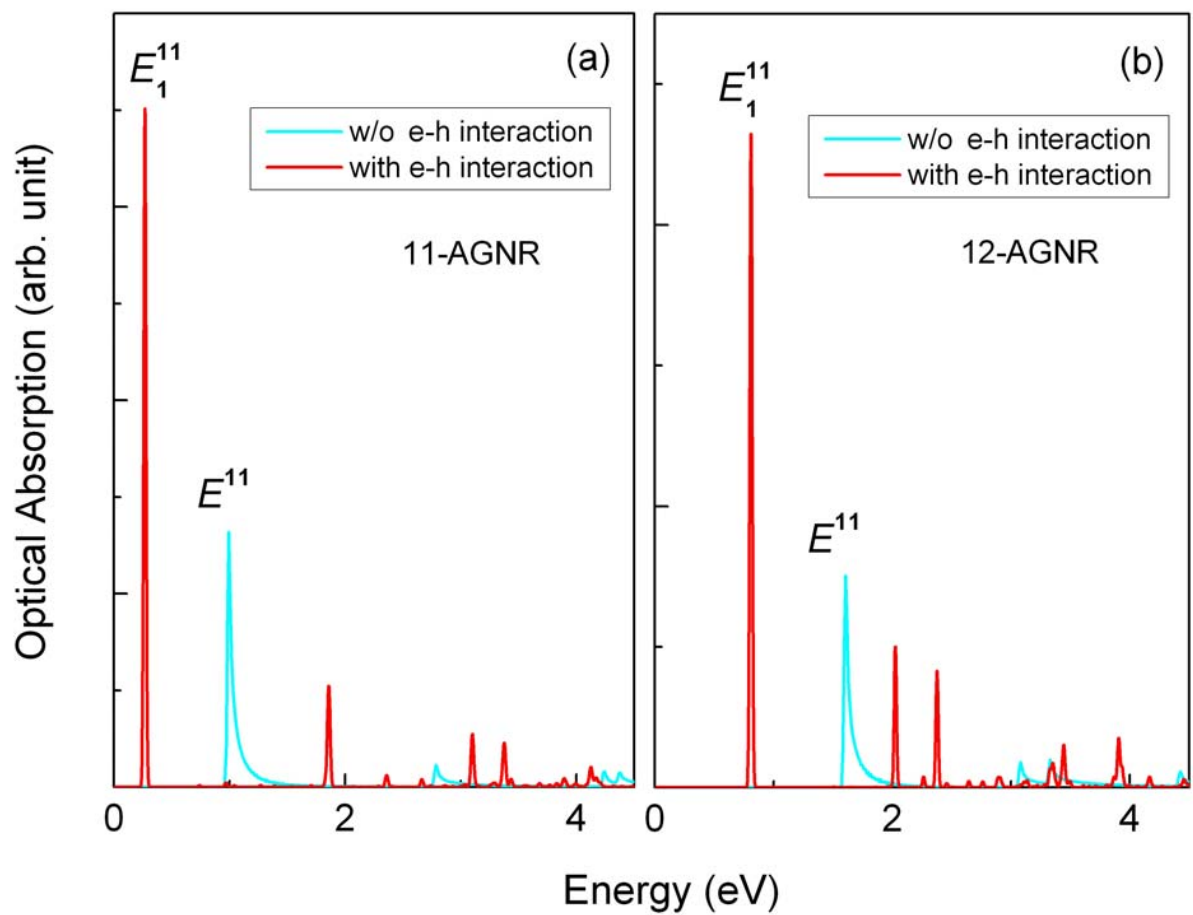

**Figure 3**. Optical spectra of 11- and 12-AGNRs. The spectra are broadened with a gaussian smearing of width of 0.01 eV.

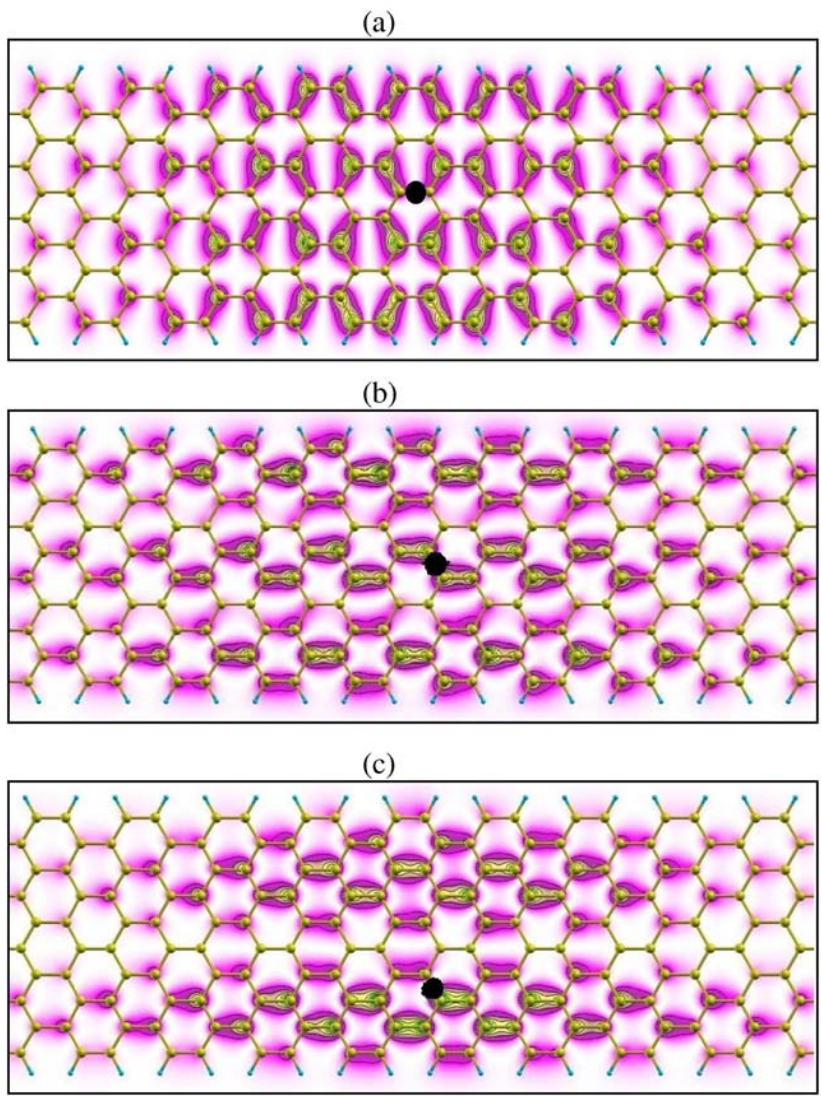

**Figure 4.** Exciton wavefunctions of the 10-AGNR for the exciton state $E_1^{11}$ (a), $E_1^{22}$ (b) and $E_1^{36}$ (c). Plotted is the electron amplitude square given the hole is fixed on top of a $\pi$ orbital at the position marked with a black spot. The distributions have been averaged along the direction perpendicular to the plane of AGNRs.

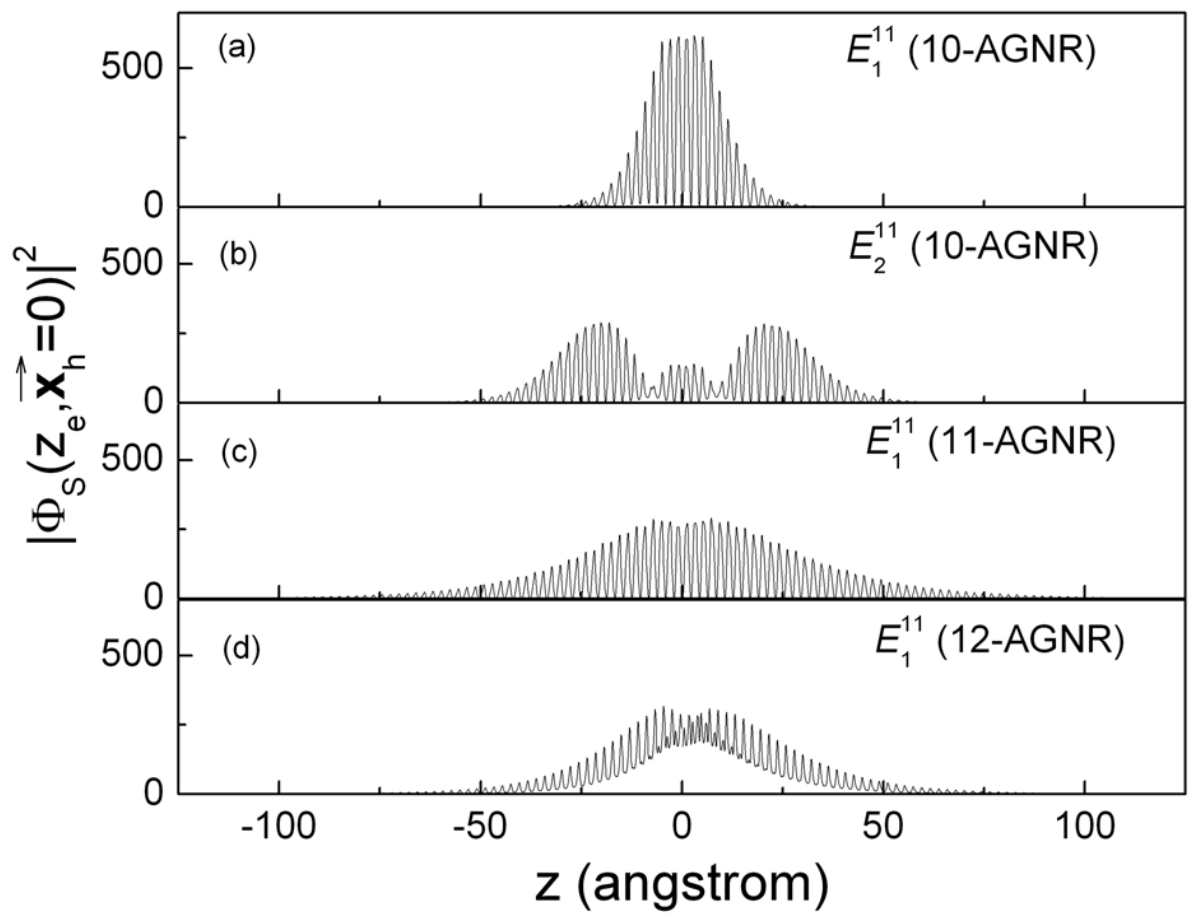

**Figure 5**. Electron distributions of excitons after integrating out coordinates perpendicular to the direction along the ribbon length. The hole is fixed on top of a π orbital which has the maximum charge density, and $z_h = 0$.